\documentclass{aa}
\input{psfig}

\def\approxgt{\mathrel{\hbox{\rlap{\lower.55ex \hbox {$\sim$}}
        \kern-.3em \raise.4ex \hbox{$>$}}}}
\def\approxlt{\mathrel{\hbox{\rlap{\lower.55ex \hbox {$\sim$}}
        \kern-.3em \raise.4ex \hbox{$<$}}}}

\begin{document}
\thesaurus{(02.01.2; 08.09.2; 08.14.1; 10.07.3; 13.25.3)}

\title{An anomalous low-state from
the globular cluster X-ray source X\thinspace1732-304 (Terzan~1)}

\author{M. Guainazzi \and A.N. Parmar \and T. Oosterbroek}

\institute{
{Astrophysics Division, Space Science Department of ESA, ESTEC, Postbus 299,
2200 AG Noordwijk, The Netherlands}
}
   
\offprints{M.Guainazzi}

\date{Received  ; accepted }

\maketitle

\markboth{M.Guainazzi et al.}{X\thinspace1732-304 (Terzan 1)}

\begin{abstract}
During a BeppoSAX observation of the X-ray burst source X\thinspace1732-304
located in the globular cluster Terzan~1 in 1999 April, the source was
detected in an anomalously low-state with an X-ray luminosity
of $1.4 \times 10^{33}$~erg~s$^{-1}$,
about 300 times fainter than the lowest
previously reported measure of the persistent flux. The spectral shape
and intensity of this low-state emission is similar to that
of a number of neutron star soft X-ray transients in quiescence, being
well fit by a $\simeq$$0.34$~keV blackbody together with a power-law with a 
photon index of $1.0 \pm^{1.0}_{1.3}$.
These similarities suggest that the same mechanism, 
such as the inhibition of
accretion due to the propeller effect, is also
operating in X\thinspace1732-304.
Alternatively, the X\thinspace1732-304
low-state may be due to obscuration of the line of sight
to the central neutron star, with only scattered and/or reflected
X-rays being detected.
The possibility that the source detected by BeppoSAX
is not X\thinspace1732-304, but a nearby dim source of the cluster field
(what would make the observed dimming of X\thinspace1732-304 even more
remarkable) is also discussed.

\end{abstract}


\section{Introduction}

Luminous (${\rm L_X \approxgt 10^{36}}$~erg~s$^{-1}$)
X-ray sources have been observed in 12 Galactic globular clusters
(see Hut et al. 1992). At least
five of them are transient sources (Verbunt et al. 1995) and
five of them have probable optical counterparts (see Deutsch et al. 1998a). 
The detection
of type I bursts from all of them but one, suggests that these
are binary systems containing neutron stars (NS). The population density
of X-ray binaries is a factor $\sim$10 higher
in globular clusters than in our Galaxy. Moreover,
no black hole transients
have been observed in globular clusters. This suggests
different formation or evolutionary scenarios for cluster and galactic
Low-Mass X-ray Binaries (LMXRBs) (Clark
1975; Verbunt 1988; Phinney \& Sigurdsson 1991).
Nevertheless, there is no compelling evidence for systematic
differences in their spectral (Christian \& Swank 1997) or timing
(Barret 1999) properties.

A bursting X-ray source was detected in the core-collapsed
(core radius ${\rm r_c = 2.75''}$; Djorgovski 1993) globular cluster Terzan~1
at the beginning of the 1980s with {\it Hakucho} (Makishima
et al. 1981; Inoue et al. 1981), and subsequently identified with the
source of soft persistent emission X\thinspace1732-304 (Parmar et al. 1989;
Johnston et al. 1995). It is now assumed that X\thinspace1732-304 is also the
source of hard X-rays observed with both SIGMA and ART-P telescopes
(Borrel et al. 1996; Pavlinsky et al. 1995).
In this {\it Paper} we report the discovery of an anomalous
low-state from X\thinspace1732-304, during which the X-ray
intensity was by a factor $\approxgt$300 lower than any previously reported
measure. 

\section{Data analysis}

The payload onboard BeppoSAX (Boella et al. 1997a) includes
two co-aligned imaging instruments: a Low-Energy Concentrator 
Spectrometer (LECS;
0.1--10~keV; Parmar et al. 1997), and a Medium-Energy 
Concentrator Spectrometer (MECS; 1.8--10~keV; Boella et al. 1997b).
The region of sky containing X\thinspace1732-304 was observed by BeppoSAX
on 1999 April 03 12:52 UT to April 04 15:29 UT.
Good data were selected from intervals when the elevation angle
above the Earth's limb was $>$$4^{\circ}$, when the instrument
configurations were nominal, and when the star tracker aligned with the 
pointing direction was operating, using the SAXDAS 2.0.0 data analysis package.
The resulting exposure times are
20.2~ks and 47.1~ks for the LECS and MECS, respectively.

\begin{figure}
\begin{center}
\psfig{figure=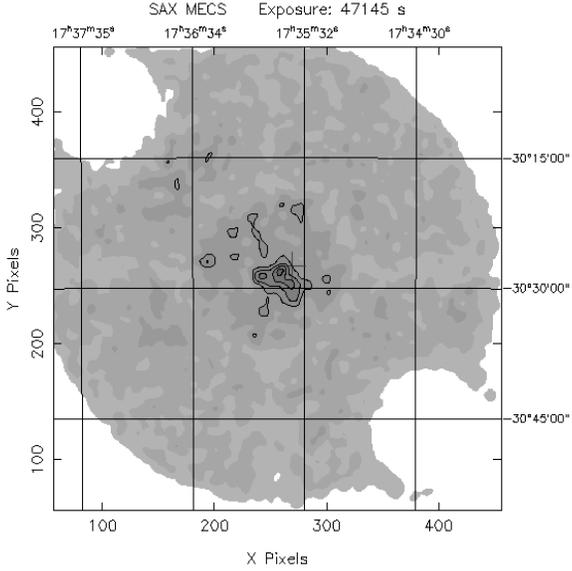,height=8.0cm,width=8.0cm,angle=-90}
\end{center}
\vspace{-0.5cm}
\caption{The MECS image of the of the X\thinspace1732-304 field. 
Solid lines indicates iso-intensity contours, in
steps of 1$\sigma$ above the local background level and starting
at 5$\sigma$. The cross indicates the HRI 
position of X\thinspace1732-304 from Johnston et
al. (1995)} 
\label{fig1}
\end{figure}

Fig.~\ref{fig1} shows the MECS image of the X\thinspace1732-304 field.
A source is detected in
the central 8$'$ region, well within the window support ring 
(where instrumental vignetting
is small and spilling of the calibration source
photons negligible). The source J2000 centroid is at
${\rm RA = 17^h\, 35^m\,43\fs7}$ and
${\rm Decl. = -30^{\circ}\, 28'\,38''}$.
The distance from the best-fit
ROSAT High Resolution Imager (HRI) centroid position
for X\thinspace1732-304
(which coincides within 5$''$ with the cluster core;
Johnston et al. 1995)
is $0\farcm9$, consistent with the
positional reconstruction accuracy for on-axis sources with BeppoSAX.
The mean background-subtracted
count rate, CR, is
${\rm (4.1 \pm 0.5) \times 10^{-3}}$~s$^{-1}$.
The MECS image is slightly
elongated in the NE-SW direction and
the contribution of any unresolved contaminating sources is
$\approxlt$25\% of the total flux. 
In the LECS the same source is detected with
${\rm CR = (2.5 \pm 0.7) \times 10^{-3}}$~s$^{-1}$.
X\thinspace1732-304 is not detected in the two high-energy non-imaging
instruments onboard BeppoSAX, consistent with 
the observed faint flux level.

We cannot exclude the possibility that the detected
source is not X\thinspace1732-304. Deep ROSAT HRI imaging of several
globular clusters
revealed a population of ${\rm L_X \sim 10^{33}}$--$10^{34}$~erg~s$^{-1}$
sources (Johnston et al. 1994; Hasinger et al.
1994). We reanalyzed the ROSAT HRI observation of
Terzan~1 and found no additional sources in its 0\fdg6 field of view.
However, the 90\% confidence upper limit to the 0.1--2~keV flux of any
undetected source within 8$'$ of X\thinspace1732-304
is $1.8 \times 10^{-13}$~erg~s$^{-1}$~cm$^{-2}$, which corresponds
to a LECS count rate of $\sim 3 \times 10^{-3}$~s$^{-1}$ in the same
band, assuming the best-fit model given in Sect.~3.
We note that if the BeppoSAX source is not
X\thinspace1732-304, then its dimming is even more dramatic than
discussed here, and most of the conclusions of this {\it Paper}
apply {\it a fortiori}. This point is discussed further in
Sect.~4.3.

\section{X-ray spectrum}

Spectra were extracted centered on the
source centroid using circular regions of 2$'$ and 4$'$
radius for the LECS and MECS, respectively.
This is a non standard extraction radius for the LECS, whose response
matrix is calibrated for an 8$'$ extraction radius. However,
the derived spectral uncertainties are dominated by the
counting statistics in such a faint source.
No significant variability is observed in the MECS 2--10~keV
light curve binned in 5760~s intervals, with a 3$\sigma$ upper 
limit on the rms fractional variability of 14\%.
Since Terzan~1 is located close
to the galactic plane (${\rm b = 0.99^{\circ}}$), it is viewed against the
diffuse emission of the galactic ridge. Therefore, the standard
BeppoSAX background subtraction technique, where deep blank field exposures
obtained at high galactic latitudes are used, is inappropriate.
Instead, a LECS background spectrum was
extracted from a semi-annulus in
the same field of view as the source, using the procedure described in Parmar
et al. (1999). A MECS background spectrum was extracted from the complement
of the source extraction region within the inner 8$'$ around the optical
center. The
spectrum was then rescaled using a multiplicative energy-independent
factor to take into account mirror vignetting. Again,
the uncertainties are
dominated by counting statistics.
Both extracted spectra were rebinned to oversample the full
width half maximum of the energy resolution by
a factor 3 and to have additionally a minimum of 20 counts 
per bin to allow use of the $\chi^2$ statistic. 
Data were selected in the energy ranges
1.0--7.0~keV (LECS) and 1.8--10.0~keV (MECS).

\begin{figure}
\begin{center}
\psfig{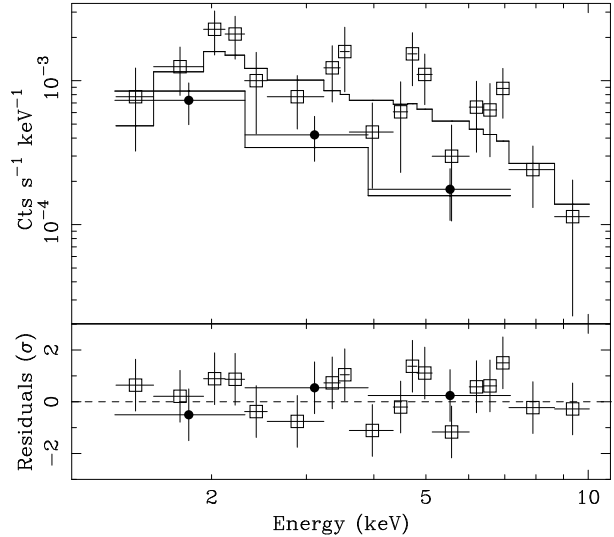}
\end{center}
\vspace{-0.5cm}
\caption{LECS (solid circles) and MECS
(empty squares) spectra. The lower panel shows the residuals in units
of standard deviations when the best-fit absorbed
blackbody and power-law model is applied to the data. Each data point
has a signal-to-noise ratio $>$1.5}
\label{fig2}
\end{figure}

Fig.~\ref{fig2} shows the LECS and MECS spectra together with the
best-fit model obtained with a simultaneous fit.
Initial fits constrained the absorbing column to be 
$<$$2.3 \times 10^{22}$~atom~cm$^{-2}$ at 90\% confidence, so this
parameter was was subsequently held fixed
at the best-fit ROSAT value of Johnston et al. (1995) of ${1.8 
\times 10^{22}}$~atom~cm$^{-2}$.
The best-fit (chance likelihood for $\chi^2$, P,
$\simeq 23\%$)
model is the combination of a blackbody with a temperature, kT,
of $0.34 \pm^{0.16}_{0.17}$~keV 
and a power-law with a photon index,
${\Gamma}$, of $1.0 \pm^{1.0}_{1.3}$ for a $\chi^2$ of
25.7 for 22 degrees of freedom (dof). 
All spectral uncertainties are quoted at 90\% confidence level for
one interesting parameter and all luminosities are for a distance
of 4.5~kpc (Ortolani et al. 1993).
The ratio between the thermal and non-thermal fluxes in the
1--10~keV energy range is 1.3.
In the simple assumption of
spherical geometry, the emission area of the blackbody
is $12 \pm^4_3$~km$^2$. 
Marginally worse fits are obtained
if power-law
(${\rm \Gamma = 2.2 \pm^{0.5}_{0.6}}$; ${\rm \chi^2 = 31.6}$ for 24~dof;
${\rm P \simeq 16\%}$)
or single
temperature bremsstrahlung models (${\rm kT = 7 \pm ^{18} _4}$~keV;
${\rm \chi^2 = 33.8}$ for 24~dof; ${\rm P \simeq 8\%}$) are employed.
The best-fit spectral parameters are given
in Table~\ref{tab1}.
An F-test indicates that the reduction in $\chi^2$
due to the addition of the blackbody to the power-law model
is significant at the 90\%
confidence level.
\begin{table}
\caption{X\thinspace1732-304
best-fit spectral parameters. In all cases the photoelectric
absorption was fixed at a value of ${\rm 1.8 \times 10^{22}}$~atom~cm$^{-2}$}
\label{tab1}
\begin{center}
\begin{tabular}{lccc} 
\hline\noalign{\smallskip}
Model              & $\Gamma$ & kT (keV) & $\chi^2$/dof \\ 
\noalign {\smallskip}
\hline\noalign {\smallskip}
Power-law (PL)        & $2.2 \pm^{0.5}_{0.6}$ & \dots & 31.6/24 \\
Bremsstrahlung     & \dots & $7 \pm^{18}_4$ & 33.8/24 \\
PL + blackbody & $1.0 \pm ^{1.0}_{1.3}$ & $0.34 \pm^{0.17}_{0.16}$ & 25.7/22 \\ 
\noalign {\smallskip}\hline 
\end{tabular}
\end{center}
\end{table}
The observed
2--10~keV flux of $(4.8 \pm 1.1) \times 10^{-13}$~erg~cm$^{-2}$~s$^{-1}$,
corresponds to an unabsorbed luminosity of 
$(1.4 \pm 0.3) \times 10^{33}$~erg~s$^{-1}$.

\section{Discussion}

\begin{figure}
\begin{center}
\hbox{
\psfig{figure=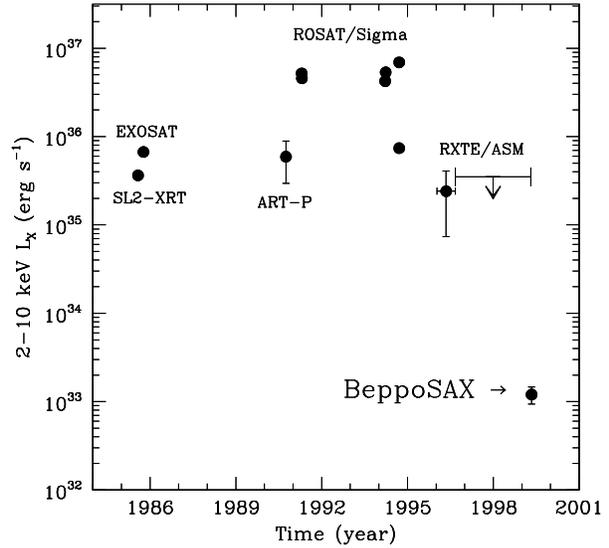,height=8.0cm,width=8.0cm,angle=0}
}
\end{center}
\vspace{-1.0cm}
\caption{The 2--10~keV historical light curve
of X\thinspace1732-304. The ordinate error bars represent the 
variability
range within each observation, except for the RXTE and
BeppoSAX observations,
where the statistical uncertainty at 90\% confidence
for one interesting parameter
is shown}
\label{fig3}
\end{figure}

Fig.~\ref{fig3} shows the historical light curve of the 2--10~keV
luminosity of X\thinspace1732-304. Data are from Borrel et al. (1996)
and references therein, except the RXTE All-Sky Monitor (ASM) 
light curve which was
retrieved from the public archive.
The uncertainties on the ASM data points include 
standard 3\% systematics. Two ASM measurements
are shown in Fig.~\ref{fig3}. This is because a 3.1$\sigma$ detection is
obtained when the first 9 months of data are accumulated and a 90\% 
confidence upper
limit is quoted for the remainder of the data.
When the reported fluxes are expressed in different
energy bands, they are converted to the 2--10~keV energy range
assuming a power-law model with
${\Gamma = 2.1}$ (Parmar et al. 1989) and
${\rm N_H = 1.8 \times 10^{22}}$~atom~cm$^{-2}$ (unless different values
are provided with the flux). The luminosity of 
X\thinspace1732-304 varied erratically
by a factor $\approxgt$20 between
$3 \times 10^{35}$ and $7 \times 10^{36}$~erg~s$^{-1}$
before the BeppoSAX observation.
BeppoSAX measured a decrease in flux of a factor $\approxgt$300,
in comparison with the previous lowest measure. The luminosity
of this low-state resembles that of NS
soft X-ray transients (SXT) in quiescence (Asai et al. 1998; 
Menou et al. 1999).

\subsection{Comparison with the historical light curves of bright GC sources}

Currently 12 Galactic globular clusters are known to contain
luminous X-ray sources. Four of these have exhibited
variations in their persistent emission by a factor $>$50. The
pattern of variability is far from homogeneous.
The sources located in NGC~6440 and NGC~6712
show occasional outbursts during which they brighten by
a factor $>$100 from a persistent luminosity of
$\sim$$10^{35}$--$10^{36}$~erg~s$^{-1}$ (Bradt \& McClintock 1983).
A recent BeppoSAX observation of an outburst from NGC~6440
(In~'t Zand et al. 1998)
revealed a decaying source with an exponential e-folding time of 6~days.
This source may also have been responsible for the 
1971 outburst (Markert et al. 1975).
This temporal variability is, however,
clearly different from that discussed here.
X\thinspace1732-304 has dimmed from an average
persistent luminosity
of $\sim$$10^{35}$--$10^{36}$~erg~s$^{-1}$ for the first time in
15~years by a factor $\approxgt$300.

The ROSAT All-Sky Survey (RASS) revealed two sources which
may show similar variability to X\thinspace1732-304.  
The source X\thinspace1745-248 in Terzan~5 was detected in the RASS
at a flux level a factor 50 higher than the {\it Einstein} measure
(Verbunt et al. 1995).
The 0.5--20~keV luminosity of its faintest state is
$1.3 \times 10^{34}$~erg~s$^{-1}$,
corresponding to a 2--10~keV luminosity of $8.8 \times 10^{33}$~erg~s$^{-1}$,
assuming the spectral shape in Verbunt et al. (1995). Similarly,
GRS\thinspace1747-312 in Terzan~6
was observed to be a factor of $\approxgt$150 brighter
than in a previous pointed ROSAT HRI observation when the
source was not detected.
Again, assuming the
spectral shape in Verbunt et al. (1995), the upper limit
to the 0.1--2.4~keV
luminosity is ${\rm < 1.9 \times 10^{34} d^2_{12}}$~erg~s$^{-1}$, where
${\rm d^2_{12}}$ is the distance in units of 12~kpc
(the distance to Terzan~6 is not well constrained,
cf. Hertz \& Wood 1985 and Djorgovski 1993). Thus the
temporal variability of the sources in Terzans~5 and 6
resembles that of X\thinspace1732-304 discussed here.
However, our discovery is remarkable in two respects: the
persistent flux variations have the highest amplitude so far
measured, and a spectroscopic measure of the ultra-dim state was
possible with BeppoSAX for the first time.

\begin{figure*}
\begin{center}
\hbox{
\hspace{0.5cm}
\psfig{figure=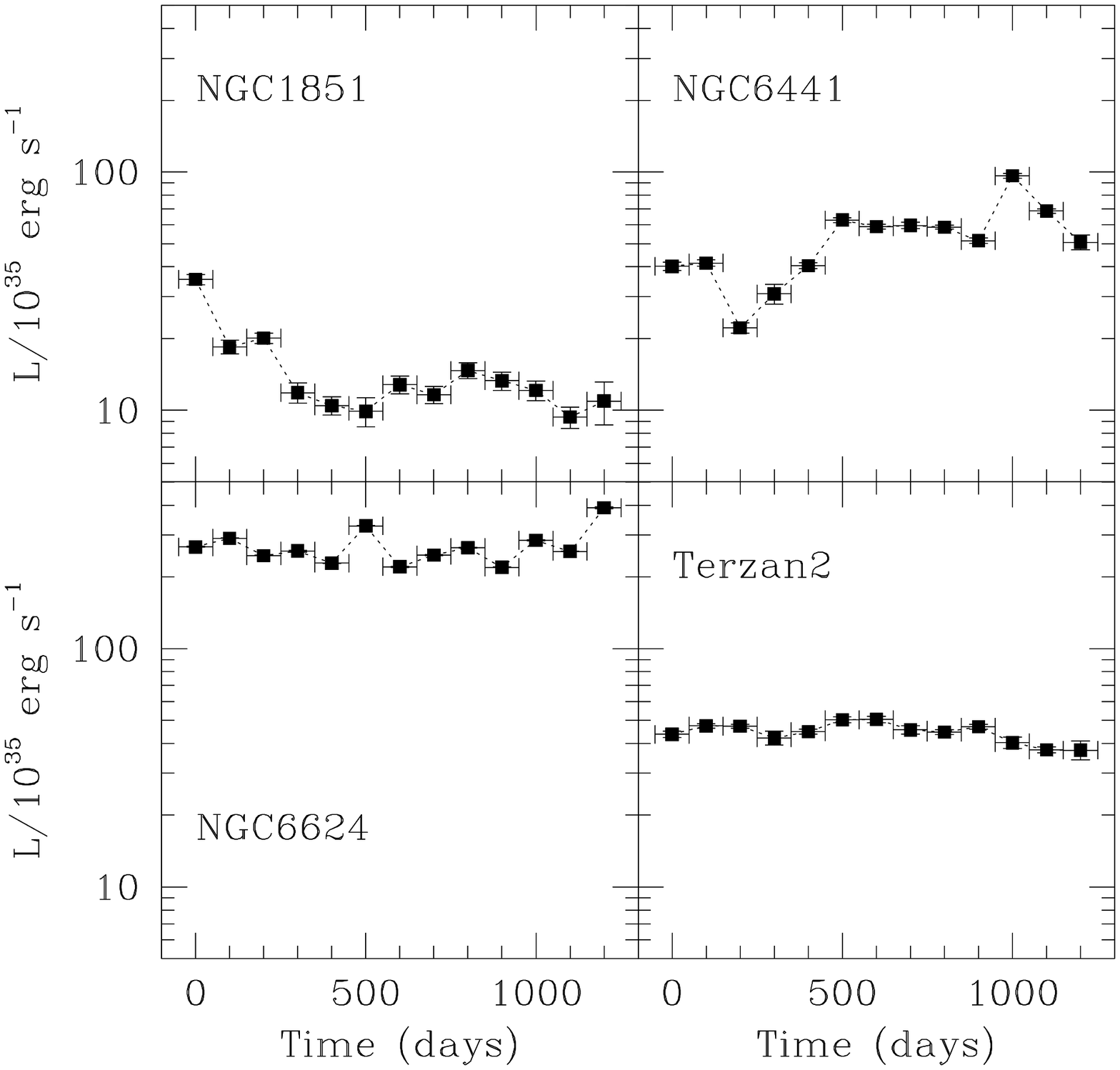,height=8.0cm,width=8.0cm,angle=0}
\hspace{0.5cm}
\psfig{figure=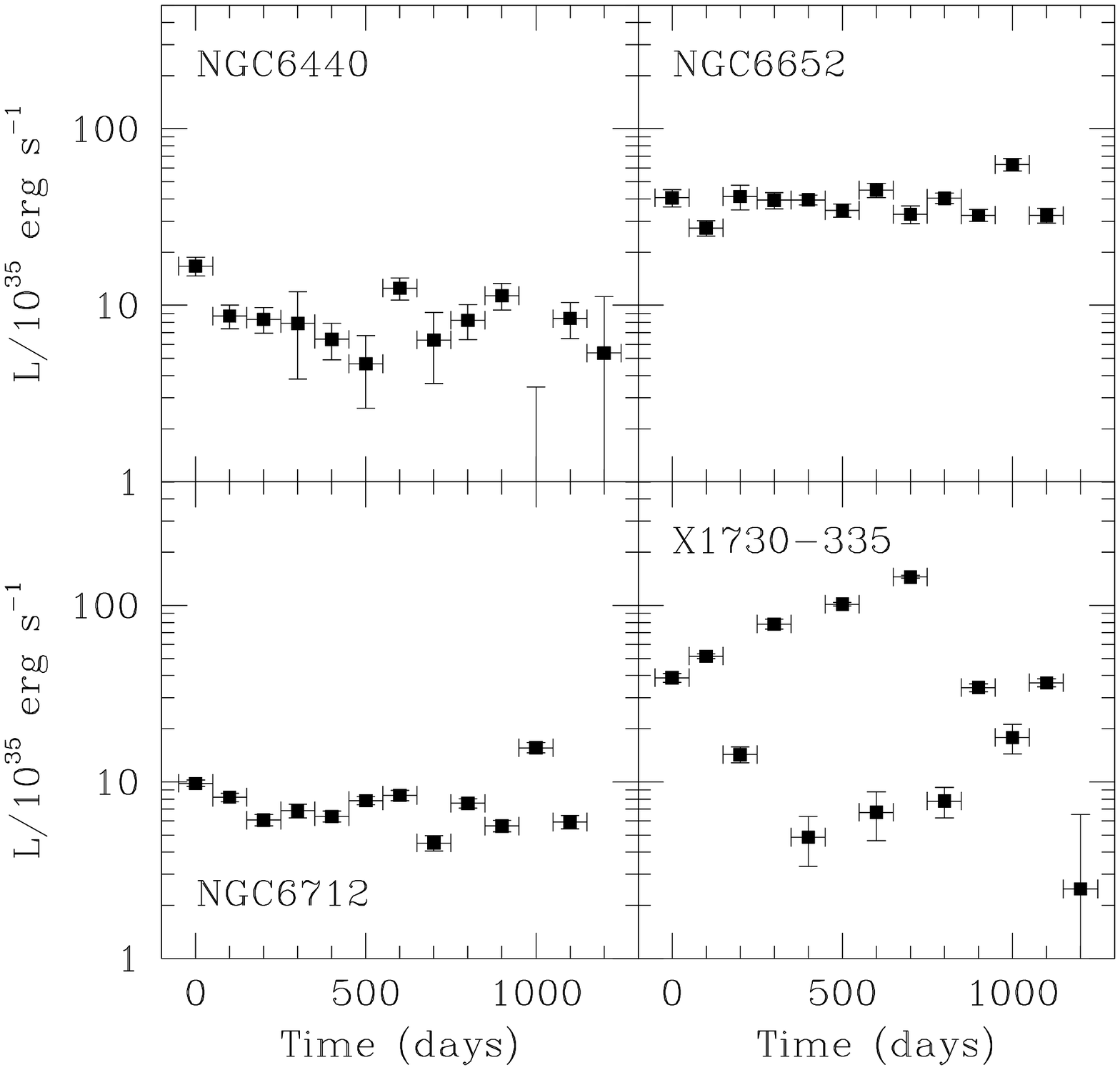,height=8.0cm,width=8.0cm,angle=0}
}
\end{center}
\vspace{-1.0cm}
\caption{Light curves of the X-ray bright globular cluster sources currently
being monitored by RXTE. The binning time is 100~days. The sources
are separated into ``persistent'' (left panels) and ``transient'' 
(right panels), according to the Verbunt et al. (1995) definition. 
X\thinspace1730-335 is the Rapid Burster in Liller~1}
\label{fig4}
\end{figure*}

Fig.~\ref{fig4} shows ASM light curves of 8 bright Galactic
globular
cluster sources which are being monitored.
RXTE is provideing the first homogeneous set of long-term light curves from
a sizeable sample of bright Galactic globular cluster X-ray sources.
The ASM count
rates have been converted to luminosities assuming the distances 
in Christian \& Swank (1997), except for Terzan~2 (Ortolani et al. 1997),
NGC\thinspace6440 (Ortolani et al. 1994) and
NGC\thinspace6652 (Djorgovski 1993). The sources are 
separated into ``persistent''
(left panels) and ``transient'' (right panels) according to the definition in
Verbunt et al. (1995). No
luminosity variations above a factor 5 are observed,
except in the 
recurrent transient X\thinspace1730-335 (the Rapid Burster) in Liller~1. This
lack of variability emphasises that the dramatic dimming 
reported here is extremely rare.

We cannot reliably estimate the timescale of the
X-ray dimming of X\thinspace1732-304 from the historical light 
curve (Fig.~\ref{fig3}).
A gradual reduction from the last 1994 ROSAT/Sigma
to the BeppoSAX observation 
cannot be excluded. However, we note that the
existence of two separate ``states'', with a very short transition time is
consistent with the ``gap'' between $10^{34.5}$~erg~s$^{-1}$ and
$10^{35.5}$~erg~s$^{-1}$ in the globular cluster X-ray 
source luminosity function
noted by Hertz \& Grindlay (1983) and Hertz \& Wood (1985).

\subsection{Possible mechanisms for the observed dimming}

The observed dramatic dimming of the X-ray luminosity in X1732-304
can be due either to a drastic reduction in
accretion rate, or to a strong obscuration of
the central X-ray source which is then only observed via
scattering and/or reflection.

Iliarionov \&
Sunyaev (1975) proposed the ``propeller effect'' as a way
of inhibiting accretion onto a magnetized compact object. This occurs when the
corotation radius ${\rm r_{co} \equiv (GM/4 \pi^2 \nu_{spin}^2)^{1/3}}$ is
smaller than the magnetospheric radius ${\rm r_m}$ (M is the NS mass)
and the centrifugal force at ${\rm r_m}$ is therefore larger than the
gravitational force. If X\thinspace1732-304 was spun-up in such a way that:
$$
{\rm \nu_{spin} > 1.7 (M/M_{\odot})^{6/7} (\dot{M}/\dot{M_E})^{3/7}_{0.1}
B_8^{-6/7} r_{NS,6}^{-18/7} \ kHz}
$$
the propeller effect would operate (King 1995). 
Here ${\rm (\dot{M}/\dot{M_E})_{0.1}}$ is the accretion
rate in units of 0.1 times Eddington, ${\rm B_8}$ the dipolar magnetic field
in units of $10^8$~Gauss and ${\rm r_{NS,6}}$ is the NS radius in units of
$10^6$~cm. The implied
kHz spin frequency is in good agreement with those observed in some
NS SXT such as Aquila~X-1 (Zhang et al. 1998)
and SAX{\thinspace}J1808.4-3658 (Wijnards \& van der Klis 1998;
Chakrabarti \& Morgan 1998).
Campana et al. (1998) interpret the change in properties
as Aquila~X-1 transitioned from outburst to quiescence 
as being due to the onset of the propeller mechanism.
Interestingly, the quiescent spectrum of Aquila~X-1 may be modeled
by the superposition of blackbody (${\rm kT \simeq 0.3}$~keV) and hard
power-law (${\Gamma = 1}$) components. Campana et al. (1998) suggest that
Aquila~X-1 could be the progenitor of a millisecond recycled
radio pulsar and they interpret the hard power-law tail as
the shock emission from the interaction between a radio pulsar wind and
the matter outflowing from the companion.

In standard thin accretion disk theory, the propeller effect
completely inhibits accretion onto the NS surface. However, low-level
X-ray emission is still present during the BeppoSAX observation. 
This can be explained if
the accretion occurs, even partly, in a spherical flow. In this case, the
centrifugal force is proportional to the polar angle ${\rm \theta}$
from the NS spin axis and hence some material can accrete close to the
polar axis. The observed blackbody emission may then
originate from the polar regions.
The fraction, f, of the NS surface that is emitting is given by:
$$
{\rm f \simeq \frac{\theta_0^2 }{2} \frac{r_{NS}}{r_{m}}}
$$
(Menou et al. 1999). Here 
${\rm \theta_0 \simeq \nu_{Keplerian}(r_m)/\nu_{spin}}$.
If ${\rm f = 0.007}$
$({\rm \equiv 12}$~km$^2$${\rm /[4 \pi (12~km)^2]}$,
see Sect.~3),
and assuming ${\rm B = 10^8}$~G,
${\rm \nu_{spin} = 10^3}$~Hz, and ${\rm M = 1.4 M_{\odot}}$, it follows
${\rm \dot{M}/\dot{M_E} \simeq 10^{-4}}$, in good agreement
with the observed low-level luminosity of X\thinspace1732-304.

Accretion can be made highly inefficient if it occurs in an
Advection Dominated flow (ADAF; Narayan \& Yi 1995; Lasota et al. 1996).
In this scenario, most of the binding energy is carried by a bulk flow
of hot plasma onto the compact object, without being substantially reradiated.
This model has been very successful in explaining the characteristics
of black hole transients in quiescence. 
In the case of NS, 
Menou et al. (1999) argue that only 0.5--10~keV luminosities
$>$$10^{34}$~erg~s$^{-1}$ can be produced  
since the infalling material impacts a hard surface where it
radiates efficiently, unless
the bulk of the matter is prevented from 
accreting onto the NS by some external mechanism (e.g.,
the propeller effect). This luminosity level is significantly
higher than that extrapolated from the BeppoSAX observation best-fit model
of X\thinspace1732-304
($<$$4 \times 10^{33}$~erg~s$^{-1}$).

Alternatively, the line of sight to the central X-ray 
source could be obscured during the low-level state,
and the residual X-rays
could originate via scattering in a hot
corona, or by reflection from some other region in the system. 
Parmar et al. (1989) 
fitted the X\thinspace1732-304 EXOSAT spectrum
with a power-law with ${\Gamma = 2.1}$
or with a 6.3~keV thermal bremsstrahlung (in the latter case with
a slightly
lower value of the absorbing column density, ${\rm N_H = 0.8 \times
10^{22}}$~atom~cm$^{-2}$). Therefore, the BeppoSAX spectrum has a similar
overall shape as observed by EXOSAT, but is a factor 400 times fainter.

The absorbing matter could be located at the radiatively bloated
inner region of the accretion disk. A ``bloated-disk'' geometry has 
been proposed to explain the low frequency QPOs observed in several LMXRBs
(Stella et al. 1987). In a different context, Nelson et al. (1997) and
van Kerkwijk et al. (1998) suggest that warping instabilities in the
accretion flow could cause the inner regions to have tilts
$>$90$^{\circ}$ during intervals of accretion torque reversal. In this case, 
the X-ray source
would be mostly observed through the accretion disk. 
This can result in column densities
as high as $\sim 10^{25} \alpha^{-4/5}$~atom~cm$^{-2}$
(van Kerkwijk et al. 1998), where ${\alpha}$ is the viscosity
parameter. Although there are several problems in applying the warped disk
model to torque reversals (cf. Sect.~3.3
in van Kerkwijk et al. 1998), it remains likely that any such warping
instabilities would be associated with enhanced absorption and increased
scattering. In this context, the low-state state of 
X\thinspace1732-304 
would result from a change in the overall 
accretion torque which caused an interval of spin-down.

It may be possible to tell which of these mechanisms is responsible 
for the low-state by studying the optical companion.
If the effects of X-ray reprocessing are still visible this would
imply that the central X-ray source is still luminous, and that our
line of sight is obscured. If no effects due to reprocessing are visible, 
then it is likely that the accretion rate is strongly reduced.
However, an optical identification is still required. 
The recent discovery of the radio counterpart to
X\thinspace1732-304 (Mart\`i et al. 1998) leaves at least four 
objects in the radio error box. Deutsch et al. (1998b)
estimate the expected range of $\lambda$336 magnitude to be
between 23 and 27, which might be accessible with the 
Wide Field Camera on board the Hubble Space Telescope.

\subsection{Could the BeppoSAX source be misidentified?}

\begin{figure}
\begin{center}
\hbox{
\psfig{figure=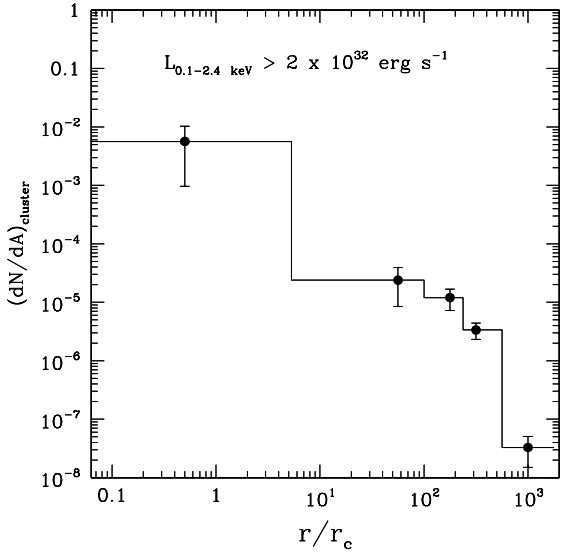,height=8.0cm,width=8.0cm,angle=0}
}
\end{center}
\vspace{-1.0cm}
\caption{Number density of X-ray sources per cluster per unit area as
a function of the distance from the cluster core (Johnston et al. 1994).
All distances
are normalized to the core radius. Only sources detected in core-collapsed
clusters with luminosities $>$$2 \times 10^{32}$~erg~s$^{-1}$ are shown}
\label{fig5}
\end{figure}

HEAO-1 (Hertz \& Wood 1985), {\it Einstein} (Hertz \& Grindlay 1983)
and ROSAT (Hasinger et al. 1994; Johnston et al. 1994) observations of
globular clusters
revealed the existence of a population of faint X-ray sources. The ROSAT
sources exhibit
0.1--2.4~keV luminosities in the range 1--6$\times 10^{32}$~erg~s$^{-1}$.
The extrapolated 0.1--2.4~keV luminosity of the source detected by BeppoSAX
is $2 \times 10^{32}$~erg~s$^{-1}$, well within the above range. It is 
therefore possible that the source detected by BeppoSAX is {\it not}
X\thinspace1732-304, but a nearby faint source.
A reanalysis of a ROSAT HRI pointed observation
of the Terzan~1 field does not
reveal any other sources close to X\thinspace1732-304.
However, a source with the same luminosity as measured by BeppoSAX could
have missed detection in this short ($\simeq$1.9~ks) ROSAT exposure.
It is difficult to quantify the probability of a chance occurrence 
of a faint source near X\thinspace1732-304.
The luminosity
distribution function is not well enough constrained 
to allow reasonable predictions for individual clusters
(Hertz \& Wood 1985). Johnston et al.
(1994) used the ROSAT Position Sensitive Proportional Counter to observe 
a small sample of nine nearby low absorption
clusters. In Fig.~\ref{fig5}
the number density of sources per cluster, normalized to the core
radius is shown. It includes only sources detected in the four
core-collapsed clusters of the Johnston et al. (1994) survey
(NGC~6397, NGC~6544, NGC~6752, NGC~7099), whose luminosity
is $>$$2 \times 10^{32}$~erg~s$^{-1}$. The function can
be approximated by a power-law of the shape: ${\rm
10^\beta (r/r_c)^{\alpha}}$, with ${\rm \alpha = 1.4 \pm 0.5}$
and ${\rm \beta = -2.4 \pm 1.0}$. This implies that the expected number of
sources within 1$'$ (2$'$) of the center of 
Terzan~1 at this luminosity threshold is 0.26 (0.40).

We note that the
Johnston et al. (1994) survey is incomplete,
and its sensitivity limit (typically
a few $10^{-14}$~erg~cm$^{-2}$~s$^{-1}$),
strongly depends on the intervening absorption. It is also unclear
if the spatial distribution of faint X-ray sources is
homogeneous from cluster to cluster, when normalized
to the core radius. There is still no clear 
understanding of the nature of these
faint sources (see Sect.~4.2 of
Johnston et al. 1994) and therefore no reason to suppose
that they come from a homogeneous
population. Nevertheless, the hypothesis that
the source detected by BeppoSAX is not X\thinspace1732-304 
cannot be excluded. If this hypothesis
is true, the observed dimming of X\thinspace1732-304 is
even more remarkable. This would not affect most of the conclusions
of this {\it paper}, except that no special mechanism would be required
to produce the observed residual flux.

\begin{acknowledgements}
  
The BeppoSAX satellite is a joint Italian-Dutch program.
MG acknowledges an ESA Research Fellowship. We acknowledge 
quick-look results provided by the ASM/RXTE team. This research
have made use of data obtained through the High Energy Astrophysics
Science Archive Research Center on-line Service, provided by the NASA/
Goddard Space Flight Center.

\end{acknowledgements}


\begin{thebibliography}{}

\bibitem{} Asai K., Dotani T., Hosi R., et al., 1998, PASJ 50, 611

\bibitem{} Barret D., 1999, ApJ, submitted

\bibitem{} Boella G., Chiappetti L., Conti G., et al., 1997b, A\&AS 122, 372

\bibitem{} Boella G., Perola G.C., Scarsi L., 1997a, A\&AS 122, 299

\bibitem{} Borrel V., Bouchet L., Jourdain E., et al., 1996, ApJ 462, 754

\bibitem{} Bradt H.V.D., McClintock J.E., 1983, ARA\&A 21, 13

\bibitem{} Campana S., Stella L., Mereghetti S., et al., 1998, ApJ 499, L65

\bibitem{} Chakrabarti D., Morgan E.H., 1998, Nat 394, 346

\bibitem{} Christian D.M., Swank J., 1997, ApJS 109, 177

\bibitem{} Clark G.W., 1975, ApJ 199, L143

\bibitem{} Deutsch E.W., Anderson S.F., Margon B., Downes R.A., 1998a, ApJ
493, 775

\bibitem{} Deutsch E.W., Margon B., Anderson S.F., 1998b, AJ 116, 1301

\bibitem{} Djorgovski S., 1993, In: ``structure and Dynamics of Globular 
Clusters'', Djorgovski S., Meylan G. (eds.), ASP Conf. Ser. 50, p.~373

\bibitem{} Hasinger G., Johnston H.M., Verbunt F., 1994, A\&A 288, 466

\bibitem{} Hertz P., Grindlay J., 1983, ApJ 275, 105

\bibitem{} Hertz P., Wood K.S., 1985, ApJ 290, 171

\bibitem{} Hut P., McMillan S., Goodman J., et al., 1992, PASP 104, 681

\bibitem{} Iliarionov A.F., Sunyaev R.A., 1975, A\&A 39, 185

\bibitem{} In~'t Zand J.J.M., Verbunt F., Strohmayer T.E., et al., 1999, A\&A 345, 100

\bibitem{} Inoue H., Koyama K., Makishima K., et al., 1981, ApJ 250, L71

\bibitem{} King A., 1995, In: Lewin W.H.G., van Paradijs J.,  
van den Heuvel E.P.J., (eds.) X-ray Binaries. Cambridge University 
Press, Cambridge, p.~419

\bibitem{} Johnston H.M., Verbunt F., Hasinger G., 1994, A\&A 289, 763

\bibitem{} Johnston H.M., Verbunt F., Hasinger G., 1995, A\&A 298, L21

\bibitem{} Lasota J.-P., Narayan R., Yi I., 1996, A\&A 314, 813

\bibitem{} Markert T.H., Backman D.E., Canizares C.R., Clark G.W., 
Levine A.M., 1975, Nat 257, 32

\bibitem{} Makishima K., Ohashi T., Inoue H., et al., 1981, ApJ 247, L23

\bibitem{} Mart\`i J., Mirabel I.F., Rodr\`iguez L.F., Chaty S., 1998, 
A\&A 332, L45

\bibitem{} Menou K., Esin A.A., Narayan R., et al., 1999, ApJ, in 
press (astroph/9810323)

\bibitem{} Narayan R., Yi I., 1995, ApJ 425, 710

\bibitem{} Nelson R.W., Bildsten L. Chakrabarti D., et al., 1997, ApJ 488, L177

\bibitem{} Ortolani S., Bica E., Barbuy B., 1993, A\&A 267, 66

\bibitem{} Ortolani S., Bica E., Barbuy B., 1994, A\&AS 108, 653

\bibitem{} Ortolani S., Bica E., Barbuy B., 1997, A\&A 326, 614

\bibitem{} Parmar A.N., Martin D.D.E., Bavdaz M., et al., 1997, A\&AS, 122, 309

\bibitem{} Parmar A.N., Oosterbroek T., Orr A., et al., 1999, A\&AS 136, 407

\bibitem{} Parmar A.N., Stella L., Giommi P., 1989, A\&A 222, 96

\bibitem{} Pavlinsky M., Grebenev S., Finogenov A., Sunyaev R., 1995, 
Advances in Space Res. 15, 95

\bibitem{} Phinney E.S., Sigurdsson S., 1991, Nat 349, 220

\bibitem{} Stella L., Priedhorsky W., White N.E., 1987, ApJ 312, L17

\bibitem{} Van Kerkwijk M.H., Chakrabarti D., Pringle J.E., Wijers R.A.M.J., 
1998, ApJ 499, L27

\bibitem{} Verbunt F., 1988, In: White N.E., Filipov L. (eds.)
 ``The Physics of Compact Objects: Theory vs. Observations'', 
Pergamon Press, Oxford, p.~529

\bibitem{} Verbunt F., Bunk W., Hasinger G., Johnston H.M., 1995, A\&A 300, 732

\bibitem{} Wijnard R., van der Klis M., 1998, Nat 394, 344

\bibitem{} Zhang S.N., Jahoda K., Kelley R.L., et al., ApJ 495, L9

\end{thebibliography}
\end{document}